\documentclass[conference]{IEEEtran}

\def\BibTeX{{\rm B\kern-.05em{\sc i\kern-.025em b}\kern-.08em
    T\kern-.1667em\lower.7ex\hbox{E}\kern-.125emX}}

\usepackage{epsfig,url}
\usepackage{amsmath,graphicx,subcaption,caption,amsthm,mathrsfs}
\usepackage{tikz,setspace}
\usepackage[nocompress]{cite}
\usetikzlibrary{decorations.pathreplacing}
\usetikzlibrary{dsp,chains,arrows,shapes,spy,fit}
\usepackage{enumitem}
\tikzset{
	font={\fontsize{9}{11.0476pt}\selectfont}}
\usepackage{mathtools,amssymb,cuted,bm}
\usepackage{pgfplots}
\pgfplotsset{compat=newest}
\usepackage{romannum}
\usepackage{algorithm,algcompatible}

\begin{document}

\title{\LARGE Study of Delay-Calibrated Joint User Activity Detection, Channel Estimation and Data Detection for Asynchronous mMTC Systems}

\author{\IEEEauthorblockN{Zhichao~Shao\textsuperscript{1}, Xiaojun Yuan\textsuperscript{2} and Rodrigo C. de Lamare\textsuperscript{3}}
	\IEEEauthorblockA{\textsuperscript{1}Yangtze Delta Region Institute (Quzhou), University of Electronic Sci. and Tech. of China, Quzhou, China}
	\IEEEauthorblockA{\textsuperscript{2}National Key Lab. of Wireless Communi., University of Electronic Sci. and Tech. of China, Chengdu, China}
	\IEEEauthorblockA{\textsuperscript{3}Department of Electrical Engineering (DEE), Pontifical Catholic University of Rio de Janeiro, Rio de Janeiro, Brazil}
	\IEEEauthorblockA{Email: \textsuperscript{1}zhichao.shao@csj.uestc.edu.cn, \textsuperscript{2}xjyuan@uestc.edu.cn, \textsuperscript{3}delamare@puc-rio.br\vspace{-0.2cm}}
}

\maketitle

\begin{abstract}
This work considers uplink asynchronous massive
machine-type communications, where a large number of low-power and low-cost devices asynchronously transmit short packets to an access point equipped with multiple receive antennas. If orthogonal preambles are employed, massive collisions will occur due to the limited number of orthogonal preambles given the preamble sequence length. To address this problem, we propose a delay-calibrated joint user activity detection, channel estimation, and data detection algorithm, and investigate the benefits of oversampling in estimating continuous-valued time delays at the receiver. The proposed algorithm is based on the expectation-maximization method, which alternately estimates the delays and detects
active users and their channels and data by noting that the collided users have different delays. Under the Bayesian inference framework, we develop a computationally efficient iterative algorithm using the approximate message passing principle to resolve the joint user activity detection, channel estimation, and data detection problem. Numerical results demonstrate the effectiveness of the proposed algorithm in terms of the normalized mean-squared errors of channel and data symbols, and the probability of misdetection.
\end{abstract}


\section{Introduction}
Massive machine-type communication (mMTC) is one of the three main use case categories for 5G networks by the 3rd Generation Partnership Project \cite{9205230}. In mMTC \cite{detmtc}, an access point (AP) is required to provide connectivity to a huge number of low-power and low-cost devices, but their activity pattern is sporadic in the uplink \cite{6525600}. Devices are kept in a sleep mode to save energy, and only a random subset of devices are activated and transmit short packets to the AP when triggered by external events \cite{7565189}. For these reasons, the development of random access (RA) design for mMTC has garnered increasing attention in recent years.

Grant-free (GF) RA \cite{9537931,8454392,listmtc}, which allows active users to send data without waiting for the grant from the AP, is particularly attractive for mMTC. In GF RA, users employ non-orthogonal preambles to enhance the access capability. The main challenge for the receiver is to jointly detect active users, estimate their channels, and recover the transmitted data. The authors in \cite{9140386} proposed a bilinear generalized approximate message passing (BiG-AMP) based receiver design to solve the joint detection and estimation problem. The authors in \cite{9103622} proposed a Bayesian receiver design, where belief propagation and expectation propagation are combined for user activity detection and channel estimation, and mean field based message passing is used for data detection. Moreover, the joint user activity detection and channel estimation problem has been studied in \cite{9714456,9691883,9390399}, where various compressive sensing-based algorithms were proposed.

Before transmitting packets, the users in the aforementioned works are required to coordinate with the AP to mitigate the impact of different transmission time delays. For example, the authors in \cite{9140386,9103622} assumed that the time delays are known and compensated in advance. The authors in \cite{msgamp,9714456,9691883,9390399} assumed that the time delays are unknown but integer multiples of the symbol duration. However, the devices in mMTC are typically low-power and low-cost, and transmit packets only when triggered by external events. As a result, coordination between the devices and the AP prior to packets transmission is difficult to achieve. 

This work considers a more practical scenario in which no prior coordination is achieved, so that time delays are unknown and continuous-valued. When triggered by external events, each active device attempts to access the AP by transmitting a randomly selected preamble from a shared pool of orthogonal preamble sequences, followed by data transmission, in an asynchronous manner. Due to the limited number of orthogonal preamble sequences, multiple active devices may select the same preamble sequence. The receiver's task is to jointly detect active users, estimate their time delays and channels, and detect data. We propose a novel delay-calibrated joint user activity detection, channel estimation, and data detection (JUCED) algorithm, and investigate the benefits of oversampling in estimating continuous-valued time delays at the receiver. Numerical results show that the proposed JUCED algorithm can achieve performance improvements 70\% greater than the baseline approach.

\section{System Model and Problem Formulation}
In this section, we present the asynchronous transmission model for mMTC and formulate the joint user activity detection, channel estimation, and data detection problem.
\subsection{Asynchronous mMTC}
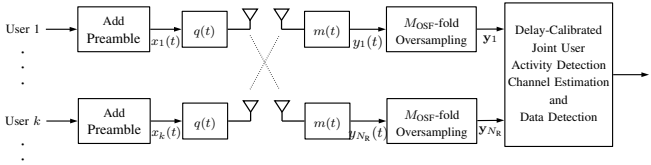
\begin{figure}[!htbp]
	\centering
	\resizebox{\columnwidth}{!}{\def\antenna{
	-- +(0mm,2.0mm) -- +(1.625mm,4.5mm) -- +(-1.625mm,4.5mm) -- +(0mm,2.0mm)
}
\tikzset{%
	harddecision/.style={draw, 
		path picture={
			\pgfpointdiff{\pgfpointanchor{path picture bounding box}{north east}}%
			{\pgfpointanchor{path picture bounding box}{south west}}
			\pgfgetlastxy\x\y
			\tikzset{x=\x*.4, y=\y*.4}
			%
			\draw (-0.5,-0.5)--(0,-0.5)--(0,0.5)--(0.5,0.5);  
			\draw (-0.25,0)--(0.25,0);
	}}
}

\begin{tikzpicture}
	\node (c0) {\footnotesize User 1};
	\node[dspsquare, right= 0.7cm of c0,minimum width=1.6cm,minimum height=1cm,text height=2em]       (c2) {\footnotesize Add \\Preamble};
	\node[dspsquare, right= 0.7cm of c2,minimum width=1cm]       (c10) {\footnotesize $q(t)$};
	\node[coordinate,right= 0.5cm of c10] (c3) {};

	\node[below= 0.15cm of c0] (c222) {\tiny \textbullet};
	\node[below= 0.01cm of c222] (c2222) {\tiny \textbullet};
	\node[below= 0.01cm of c2222] (c22222) {\tiny \textbullet};
	\node[coordinate,below= 0.5cm of c2] (cfix) {};
	
	\node[below= 1.6cm of c0] (c00) {\footnotesize User $k$};
	\node[dspsquare, right= 0.7cm of c00,minimum width=1.6cm,minimum height=1cm,text height=2em]                    (c22) {\footnotesize Add\\ Preamble};
	\node[dspsquare, right= 0.7cm of c22,minimum width=1cm]       (c12) {\footnotesize $q(t)$};
	\node[coordinate,right= 0.5cm of c12] (c33) {};
	
	\node[below= 0.15cm of c00] (c2225) {\tiny \textbullet};
	\node[below= 0.01cm of c2225] (c22225) {\tiny \textbullet};
	
	\node[coordinate,right= 0.7cm of c3] (c8) {};
	\node[coordinate,right= 0.7cm of c33] (c9) {};
	
	\node[coordinate,right= 0.5cm of c10.-15] (c88) {};
	\node[coordinate,below= 1.2cm of c88] (c99) {};
	
	\node[dspsquare, right= 0.5cm of c8,minimum width=1cm]       (c42) {\footnotesize $m(t)$};
	\node[dspsquare, right= 0.5cm of c9,minimum width=1cm]       (c43) {\footnotesize $m(t)$};
	
	\node[dspsquare, right= 0.8cm of c42,text width=2cm,minimum width=1cm,minimum height=1cm,text height=-1em]       (c13) {\footnotesize\doublespacing $M_{\text{OSF}}$-fold \\ Oversampling};
	\node[dspsquare, right= 0.8cm of c43,text width=2cm,minimum width=0.5cm,minimum height=1cm,text height=-1em]       (c14) {\footnotesize\doublespacing $M_{\text{OSF}}$-fold \\ Oversampling};
	
	\node[dspsquare,right= 8.7cm of cfix,minimum height=3.2cm,text height=6.2em,minimum width=2.4cm] (c17) {\footnotesize Delay-Calibrated\\\footnotesize Joint User \\\footnotesize Activity Detection\\\footnotesize  Channel Estimation\\\footnotesize and\\\footnotesize Data Detection};
	
	\node[coordinate,right= 0.65cm of c88] (c200) {};
	\node[coordinate,right= 0.65cm of c99] (c211) {};
	
	\node[coordinate,right= 0.8cm of c17] (c40) {};
	
	
	\draw[dspconn] (c0) -- node[midway,below] {} (c2);
	\draw[dspconn] (c2) -- node[midway,below] {\footnotesize $x_1(t)$} (c10);
	\draw[thick] (c10) -- node[] {} (c3);
	\draw[dspconn] (c00) -- node[midway,below] {} (c22);
	\draw[dspconn] (c22) -- node[midway,below] {\footnotesize $x_k(t)$} (c12);
	\draw[thick] (c12) -- node[] {} (c33);
	\draw [thick] (c3) \antenna;
	\draw [thick] (c33) \antenna;
	\draw [thick] (c8) \antenna;
	\draw [thick] (c9) \antenna;
	\draw[densely dotted] (c88) -- node[] {} (c211);
	\draw[densely dotted] (c99) -- node[] {} (c200);
	\draw[thick] (c8) -- node[] {} (c42);
	\draw[thick] (c9) -- node[] {} (c43);
	\draw[dspconn] (c42) -- node[midway,below] {$y_1(t)$} (c13);
	\draw[dspconn] (c43) -- node[midway,below] {$y_{N_\text{R}}(t)$} (c14);
	\draw[dspconn] (c13) -- node[midway,below] {\footnotesize $\mathbf{y}_1$} (c17.140);
	\draw[dspconn] (c14) -- node[midway,below] {\footnotesize $\mathbf{y}_{N_\text{R}}$} (c17.-140);
	\draw[dspconn] (c17) -- node[midway,above] {} (c40);
	
\end{tikzpicture}}
	\caption{System model for the scenario of asynchronous mMTC.}
	\label{fig:system_model}
\end{figure}

Consider a wireless communication system consisting of one AP equipped with $N_\text{R}$ receive antennas and several single-antenna users, as shown in Fig. \ref{fig:system_model}. While accessing the AP, each active user transmits one packet with $N_\text{F}$ symbols. Then the signal transmitted by user $k$ can be expressed by
\begin{equation}
	x_{k}(t) = x_{k0}\delta(t)+\cdots+x_{k(N_{\text{F}}-1)}\delta(t-(N_{\text{F}}-1)T_\text{S}),
	\label{equ_x_S}
\end{equation}
where $x_{kn}$ denotes the symbol sent by user $k$ at time instant $nT_\text{S}$, $T_\text{S}$ is the symbol interval. In each packet, the first $N_\text{P}$ symbols are preamble sequences, and the remaining $N_\text{D}=N_\text{F}-N_\text{P}$ symbols are data.
After going through the pulse shaping filter, denoted by $q(t)$, the signal is transmitted over the channel. 

At the AP, the messages arriving at the $n_\text{R}$-th receive antenna are written as
\begin{equation}\label{equ_conv}
	s_{n_\text{R}}(t)=\sum_kh_{kn_\text{R}}z(t)\ast x_{k}(t-\tau_k),
\end{equation}
where $h_{kn_\text{R}}(t)=h_{kn_\text{R}}\delta(t)$ is the impulse response of the channel from user $k$ to the $n_\text{R}$-th receive antenna with $h_{kn_\text{R}}$ being the complex gain, $z(t)$ is the convolution of $q(t)$ and $m(t)$, $\tau_k$ is the time delay of the transmission from user $k$ to the AP. Note that $\tau_k$ may not be an integer multiple of $T_\text{S}$ due to the very limited coordination between users and the AP. 

Considering the environmental noise, denoted by $w_{n_\text{R}}(t)$, the overall received signal at the $n_\text{R}$-th antenna of the AP is
\begin{equation}\label{sysoverall}
		y_{n_\text{R}}(t)=\sum_kh_{kn_\text{R}}z(t)\ast x_{k}(t-\tau_k)+m(t)\ast w_{n_\text{R}}(t).
\end{equation}
The continuous signal $y_{n_\text{R}}(t)$ then flows through the oversampling module to yield the discrete samples with a sampling interval of $\frac{T_\text{S}}{M_{\text{OSF}}}$, where $M_{\text{OSF}}$ denotes the oversampling factor.

With the asynchronous transmission, the received signal is frameless. The sliding window technique \cite{shao2024} can be applied to continuously detect active users and recover transmitted packets. In the following, we focus on one observation window with a width of $N_\text{W}T_\text{S}$ and a start time of $t_\text{W}$, where $N_\text{W}$ is chosen larger than $N_\text{F}$. Assuming that there are $K$ users in the considered window and after the oversampling module, 
	the discrete samples of $y_{n_\text{R}}(t)$ in \eqref{sysoverall} are
	\begin{align}\label{discretnoise}
		\nonumber&y_{n_\text{R}}\left[\frac{i}{M_{\text{OSF}}}\right]=\sum_{j=0}^{N_\text{W}M_{\text{OSF}}-1}\sum_{k=1}^{K}\alpha_{k}h_{kn_\text{R}}z\left[\frac{i-j}{M_{\text{OSF}}}\right] x_{k}\left[\frac{j}{M_{\text{OSF}}}\right]\\&\hspace{-0.2cm}+m\left[\frac{i-j}{M_{\text{OSF}}}\right]w_{n_\text{R}}\left[\frac{j}{M_{\text{OSF}}}\right], 
		i\in\{0,\cdots,N_\text{W}M_{\text{OSF}}-1\},
	\end{align}
	where $y_{n_\text{R}}[\frac{i}{M_{\text{OSF}}}]=y_{n_\text{R}}(t)|_{t=t_\text{W}+\frac{i}{M_{\text{OSF}}}T_\text{S}}$,  $z[\frac{i-j}{M_{\text{OSF}}}]=z(t)|_{t=\frac{i-j}{M_{\text{OSF}}}T_\text{S}}$, and 
	\begin{equation}
		\hspace{-0.2cm}		x_{k}\left[\frac{j}{M_{\text{OSF}}}\right]=
		\begin{cases}
			x_{kn}, \text{ if }\lfloor M_{\text{OSF}}\frac{t_\text{W}-\tau_{k}}{T_\text{S}}\rfloor+j=nM_{\text{OSF}},\\\qquad n\in \{0,\cdots,N_\text{F}-1\},\\
			0,\text{ \quad otherwise},
		\end{cases}
	\end{equation}
	with $\lfloor\cdot\rfloor$ denoting rounding to the nearest integer less than or equal to the input variable. The variable $\alpha_{k}$ is the binary user activity indicator. The vector form of \eqref{discretnoise} is described by
	\begin{equation}
		\begin{split}
			\mathbf{y}_{n_\text{R}}=\sum_{k=1}^{K}\alpha_{k}h_{kn_\text{R}}\mathbf{Z}\mathbf{x}_{k}(\tau_{k})+\mathbf{F}\mathbf{w}_{n_\text{R}},
		\end{split}
		\label{equ_siso}
	\end{equation}
	where  $\mathbf{y}_{n_\text{R}}=[0,\cdots,y_{n_\text{R}}[N_\text{W}-\frac{1}{M_{\text{OSF}}}]]^T\in\mathbb{C}^{N_\text{W}M_{\text{OSF}}\times1}$, $\mathbf{x}_{k}=[0,\cdots,x_{k}[N_\text{W}-\frac{1}{M_{\text{OSF}}}]]^T\in\mathbb{C}^{N_\text{W}M_{\text{OSF}}\times1}$. The vector $\mathbf{w}_{n_\text{R}}\sim \mathcal{CN}(\mathbf{0}_{N_\text{W}M_{\text{OSF}}\times1},\sigma_\text{W}^2\mathbf{I}_{N_\text{W}M_{\text{OSF}}})$ represents the complex Gaussian noise samples. The matrix $\mathbf{Z}\in\mathbb{R}^{N_\text{W}M_{\text{OSF}} \times N_\text{W}M_{\text{OSF}}}$ is a Toeplitz matrix constructed by $z(t)$ at different time instants with the following form
	\begin{equation}
		\resizebox{\columnwidth}{!}{$\mathbf{Z} = \begin{bmatrix}
				z[0] & \dots & z[3] & 0 & 0 & \dots & 0\\ 
				z[-\frac{1}{M_{\text{OSF}}}] & \dots & z[3-\frac{1}{M_{\text{OSF}}}] & z[3] & 0 & \dots & 0\\
				\vdots & \ddots & \vdots& \vdots& \vdots& \ddots& \vdots\\
				0 & \dots & 0& 0 & z[-3] & \dots & z[0]\\
			\end{bmatrix}.$}
		\label{eq_zform}
	\end{equation}
	Since most of the energy of $z(t)$ is concentrated on a finite range, we consider $z[\frac{i-j}{M_{\text{OSF}}}]$ in the range of $\frac{i-j}{M_{\text{OSF}}}\in[-3,3]$, and omit $z[\frac{i-j}{M_{\text{OSF}}}]$ out of this range. The matrix $\mathbf{F}\in\mathbb{R}^{N_\text{W}M_{\text{OSF}} \times N_\text{W}M_{\text{OSF}}}$ is a Toeplitz matrix constructed by $m(t)$ at different time instants with the form similar to \eqref{eq_zform}. Similar structures have been used in \cite{jidf,dynovs} for interpolation, decimation and oversampling.
	
	Considering all the receive antennas, \eqref{equ_siso} is extended as 
	\begin{equation}\label{equ_sys1}
		\mathbf{Y}=[\mathbf{y}_{1},\cdots,\mathbf{y}_{N_\text{R}}]=
		\mathbf{Z}\mathbf{X}(\bm{\tau})\mathbf{G} + \mathbf{FW},
	\end{equation}
	where $\mathbf{Y}\in\mathbb{C}^{N_\text{W}M_{\text{OSF}}\times N_\text{R}}$ contains the received samples, $\mathbf{X}(\bm{\tau})=[\mathbf{x}_{1}(\tau_{1}),\cdots,\mathbf{x}_{K}(\tau_{K})]\in\mathbb{C}^{N_\text{W}M_{\text{OSF}}\times K}$ contains the delayed packets with time delays $\bm{\tau}=[\tau_{1},\cdots,\tau_{K}]^T\in\mathbb{R}^{K\times 1}$,   $\mathbf{W}=[\mathbf{w}_{1},\cdots,\mathbf{w}_{N_\text{R}}]\in\mathbb{C}^{N_\text{W}M_{\text{OSF}}\times N_\text{R}}$ contains the noise samples. The matrix $\mathbf{G}\in\mathbb{C}^{K\times N_\text{R}}$ is a row sparse equivalent channel matrix with non-zero rows representing active users.  
	
	\subsection{Problem Formulation}
	The task is to estimate $\bm{\tau}$, $\mathbf{G}$ and $\mathbf{X}$ from the noisy observation $\mathbf{Y}$ in \eqref{equ_sys1}. The maximum likelihood estimate of $\bm{\tau}$ is given as
	\begin{equation}\label{equ_ml}
		\hat{\bm{\tau}}=\arg\max_{\bm{\tau}}\ln\int_{\mathbf{X},\mathbf{G}} p(\mathbf{X},\mathbf{G},\mathbf{Y};\bm{\tau}),
	\end{equation}
	where $p(\mathbf{X},\mathbf{G},\mathbf{Y};\bm{\tau})$ denotes the joint density of $\mathbf{X}$, $\mathbf{G}$ and $\mathbf{Y}$ with the variable $\bm{\tau}$. It is not a trivial task to solve the above problem since both the data parts of $\mathbf{X}$ and $\mathbf{G}$ are unknown. According to the evidence lower bound, the log-likelihood function in \eqref{equ_ml} is lower bounded by
	\begin{align}\label{equ_mlpro}
		\nonumber&\ln\int_{\mathbf{X},\mathbf{G} }p(\mathbf{X},\mathbf{G},\mathbf{Y};\bm{\tau})=\ln E_{p'(\mathbf{X},\mathbf{G};\bm{\tau})}\left\{\frac{ p(\mathbf{X},\mathbf{G},\mathbf{Y};\bm{\tau})}{p'(\mathbf{X},\mathbf{G};\bm{\tau})}\right\}\\&\hspace{2.5cm}\geq E_{p'(\mathbf{X},\mathbf{G};\bm{\tau})}\left\{\ln\frac{ p(\mathbf{X},\mathbf{G},\mathbf{Y};\bm{\tau})}{p'(\mathbf{X},\mathbf{G};\bm{\tau})}\right\},
	\end{align}
	where $p'(\mathbf{X},\mathbf{G};\bm{\tau})$ is any distribution related to $\mathbf{X}$ and $\mathbf{G}$ with the variable $\bm{\tau}$,  $E_{p'(\mathbf{X},\mathbf{G};\bm{\tau})}\{\cdot\}$ represents the expectation over the distribution $p'(\mathbf{X},\mathbf{G};\bm{\tau})$. The equality of \eqref{equ_mlpro} holds when  $p'(\mathbf{X},\mathbf{G};\bm{\tau})=p(\mathbf{X},\mathbf{G}|\mathbf{Y};\bm{\tau})$. Thus, \eqref{equ_ml} is equivalent to 
	\begin{equation}\label{equ_em}
		\hat{\bm{\tau}}=\arg\max_{\bm{\tau}}E_{p(\mathbf{X},\mathbf{G}|\mathbf{Y};\bm{\tau})}\left\{\ln\frac{ p(\mathbf{X},\mathbf{G},\mathbf{Y};\bm{\tau})}{p(\mathbf{X},\mathbf{G}|\mathbf{Y};\bm{\tau})}\right\}.
	\end{equation}
	The posterior density $p(\mathbf{X},\mathbf{G}|\mathbf{Y};\bm{\tau})=p(\mathbf{X}|\mathbf{Y};\bm{\tau})p(\mathbf{G}|\mathbf{Y};\bm{\tau})$ is difficult to obtain since $\bm{\tau}$ is the variable to be estimated. In the next section, we propose a low-complexity EM framework \cite{em} to solve the above problem. With $\hat{\bm{\tau}}$, the minimum mean-square error (MMSE) estimators of $\mathbf{X}$ and $\mathbf{G}$ are
	\begin{align}\label{equ_mmse}
		\hat{\mathbf{X}}=E_{p(\mathbf{X}|\mathbf{Y};\hat{\bm{\tau}})}\{\mathbf{X}\}\qquad
		\hat{\mathbf{G}}=E_{p(\mathbf{G}|\mathbf{Y};\hat{\bm{\tau}})}\{\mathbf{G}\}.
	\end{align}
	
\section{Delay-Calibrated JUCED}\label{sec2}

The proposed algorithm is based on the EM framework \cite{em} that consists of two steps named the E step and the M step. In the E step, a message passing based JUCED algorithm is proposed to obtain the posterior density $p(\mathbf{X}|\mathbf{Y};\bm{\tau})$ and $p(\mathbf{G}|\mathbf{Y};\bm{\tau})$. This is in contrast with standard iterative detection and decoding schemes \cite{spa,mfsic,mbdf,bfidd,1bitidd,llrref,risidd,apsllr,iddocl} that exchange messages with receive filters, channel estimators and decoders performing their specific tasks. In the M step, a greedy search based delay-calibration method \cite{shao2024} is utilized to estimate $\bm{\tau}$. Note that since the filtered noise samples $\mathbf{F}\mathbf{W}$ in \eqref{equ_sys1} are correlated with non-diagonal covariance $\sigma_\text{W}^2\mathbf{F}\mathbf{F}^H$,the pre-whitening technique is first applied to  \eqref{equ_sys1} to facilitate the message passing in JUCED described in Section \ref{sec_delay}. 
	
	\subsection{EM Framework}\label{sec2em}
	In the $u_\text{O}$-th EM iteration, the E step calculates the expectation  $E_{p(\mathbf{X},\mathbf{G}|\mathbf{Y};\hat{\bm{\tau}}^{u_\text{O}})}\left\{\ln\frac{ p(\mathbf{X},\mathbf{G},\mathbf{Y};\bm{\tau})}{p(\mathbf{X},\mathbf{G}|\mathbf{Y};\hat{\bm{\tau}}^{u_\text{O}})}\right\}$ based on the posterior density $p(\mathbf{X},\mathbf{G}|\mathbf{Y};\hat{\bm{\tau}}^{u_\text{O}})$, and the M step computes $\hat{\bm{\tau}}^{u_\text{O}+1}$ for maximizing the expectation in the E step, where $\hat{\bm{\tau}}^{u_\text{O}}$ denotes the estimation of $\bm{\tau}$ in the $u_\text{O}$-th EM iteration.
	
	The E step in the $u_\text{O}$-th EM iteration is calculated as 
	\begin{align}\label{equ_expec} \nonumber&\hspace{0.5cm}\int_{\mathbf{X},\mathbf{G}} p(\mathbf{X},\mathbf{G}|\mathbf{Y};\hat{\bm{\tau}}^{u_\text{O}})\ln p(\mathbf{Y}|\mathbf{X},\mathbf{G};\bm{\tau})\\
		\nonumber&\hspace{1.3cm}+\int_{\mathbf{X},\mathbf{G}} p(\mathbf{X},\mathbf{G}|\mathbf{Y};\hat{\bm{\tau}}^{u_\text{O}})\ln p(\mathbf{X},\mathbf{G};\bm{\tau})\\
		\nonumber&\propto\frac{1}{\sigma_\text{W}^2}\sum_{n_\text{R}=1}^{N_\text{R}}2\text{Re}\left\{\mathbf{Y}_{n_\text{R}}^H\mathbf{Z}\hat{\mathbf{X}}^{u_\text{O}}(\bm{\tau})\hat{\mathbf{g}}_{n_\text{R}}^{u_\text{O}}\right\}-\text{Tr}\left\{\left((\mathbf{Z}\hat{\mathbf{X}}^{u_\text{O}}(\bm{\tau}))^H\right.\right.\\
		\nonumber&\hspace{0.5cm}\left.\left.\times\mathbf{Z}\hat{\mathbf{X}}^{u_\text{O}}(\bm{\tau})+\mathbf{Z}^H\mathbf{V}^{u_\text{O}}_{\mathbf{X}(\bm{\tau})}\mathbf{Z}\right)\left(\hat{\mathbf{g}}_{n_\text{R}}^{u_\text{O}}(\hat{\mathbf{g}}_{n_\text{R}}^{u_\text{O}})^H+\mathbf{V}^{u_\text{O}}_{\mathbf{g}_{n_\text{R}}}\right)\right\}\\
		&\hspace{0.5cm}+\sum_{k=1}^{K}-\text{Tr}\left\{\hat{\mathbf{x}}^{u_\text{O}}_{k}(\tau_k)\hat{\mathbf{x}}^{u_\text{O}}_{k}(\tau_k)^H+\mathbf{V}^{u_\text{O}}_{\mathbf{x}_{k}(\tau_k)}\right\},
	\end{align}
	where $p(\mathbf{X},\mathbf{G}|\mathbf{Y};\hat{\bm{\tau}}^{u_\text{O}})=p(\mathbf{X}|\mathbf{Y};\hat{\bm{\tau}}^{u_\text{O}})p(\mathbf{G}|\mathbf{Y};\hat{\bm{\tau}}^{u_\text{O}})$ is obtained by the proposed JUCED in Section \ref{sec_delay}, and the terms irrelevant to $\bm{\tau}$ are omitted. 
	
	The M step in the $u_\text{O}$-th EM iteration is calculated as 
	\begin{equation}\label{equ_maxi}
		\hat{\bm{\tau}}^{u_\text{O}+1}=\arg\max_{\bm{\tau}}f(\bm{\tau}),
	\end{equation}
	where $f(\bm{\tau})$ denotes the result of \eqref{equ_expec}.
	Obtaining a closed-form solution to \eqref{equ_maxi} is challenging due to the non-linear relationship between $\mathbf{X}$ and $\bm{\tau}$. The greedy search based delay-calibration algorithm \cite{shao2024} is utilized to obtain a sub-optimal solution to \eqref{equ_maxi}. 
	
	\subsection{JUCED}\label{sec_delay}
	We derive the approximate message flows to obtain $p(\mathbf{X}|\mathbf{Y};\hat{\bm{\tau}}^{u_\text{O}})$ and $p(\mathbf{G}|\mathbf{Y};\hat{\bm{\tau}}^{u_\text{O}})$. By defining the intermediate variables $\mathbf{A} \triangleq \mathbf{B}\mathbf{G}$ and $\mathbf{B} \triangleq \mathbf{Z}\mathbf{X}(\hat{\bm{\tau}}^{u_\text{O}})$, 
	and using Bayes’ theorem, the joint posterior distribution of $\mathbf{A}$, $\mathbf{B}$, $\mathbf{G}$ and $\mathbf{X}$ can be expressed by using the factorized distribution as
	\begin{align}\label{equ_fact}
		\nonumber p(\mathbf{A},\mathbf{B},\mathbf{G},\mathbf{X}|\mathbf{Y};\hat{\bm{\tau}}^{u_\text{O}})=&\frac{1}{p(\mathbf{Y})}p(\mathbf{Y}|\mathbf{A})p(\mathbf{A}|\mathbf{B},\mathbf{G})\\&\times p(\mathbf{G})p(\mathbf{B}|\mathbf{X})p(\mathbf{X};\hat{\bm{\tau}}^{u_\text{O}}),
	\end{align}
	where $p(\mathbf{Y})=\int_{\mathbf{A},\mathbf{B},\mathbf{G},\mathbf{X}} p(\mathbf{A},\mathbf{B},\mathbf{G},\mathbf{X},\mathbf{Y})$ is the normalization constant, 
	\begin{subequations}
		\begin{align}\label{equ_fg1}
			&p(\mathbf{Y}|\mathbf{A}) = \prod_{n_\text{R}=1}^{N_\text{R}}\prod_{n=1}^{N_\text{W}M_{\text{OSF}}}\mathcal{CN}(y_{nn_\text{R}};a_{nn_\text{R}},\sigma_\text{W}^2),
			\\\label{equ_fg2}
			&p(\mathbf{A}|\mathbf{B},\mathbf{G}) =\prod_{n_\text{R}=1}^{N_\text{R}}\prod_{n=1}^{N_\text{W}M_\text{OSF}} \delta(a_{nn_\text{R}}-\sum_{k=1}^{K}b_{nk}g_{kn_\text{R}}),
			\\\label{equ_fg3}
			&p(\mathbf{G}) =\prod_{k=1}^{K}((1-\rho)\delta(\mathbf{g}_{k})+\rho \mathcal{CN}(\bm{0}_{N_\text{R}\times1},\lambda\mathbf{I}_{N_\text{R}})),
			\\\label{equ_fg4}
			&p(\mathbf{B}|\mathbf{X}) = \prod_{k=1}^{K}\prod_{n=1}^{N_\text{W}M_{\text{OSF}}}\delta(b_{nk}-\sum_{i=1}^{N_\text{W}M_{\text{OSF}}}z_{ni}x_{ik}),
			\\\label{equ_fg5}
			\nonumber	&p(\mathbf{X};\hat{\bm{\tau}}^{u_\text{O}}) = \prod_{k=1}^{K}\prod_{i=\hat{\tau}_k^{u_\text{O}}+1}^{\hat{\tau}_k^{u_\text{O}}+N_\text{P}M_{\text{OSF}}}\delta(x_{ik}-x_{\text{P},ik})\\&\hspace{2cm}\times\prod_{i=\hat{\tau}_k^{u_\text{O}}+N_\text{P}M_{\text{OSF}}+1}^{\hat{\tau}_k^{u_\text{O}}+N_\text{F}M_{\text{OSF}}}\mathcal{CN}(x_{ik};0,1).
		\end{align}
	\end{subequations}
	In \eqref{equ_fg3}, $\rho$ is the probability of active users and $\lambda$ is the average path-loss and shadowing component. In \eqref{equ_fg5}, $x_{\text{P},ik}$ denotes the preamble symbol, and each data symbol is a Gaussian random variable with mean 0 and variance 1. 
	
	According to the sum-product rule and the factor graph, the message from node $p(a_{nn_\text{R}}|\cdot)$ to node $a_{nn_\text{R}}$ at iteration $u_\text{I}$, denoted by $\mathcal{M}^{u_\text{I}}_{p(a_{nn_\text{R}}|\cdot)\rightarrow a_{nn_\text{R}}}(a_{nn_\text{R}})$, is approximated as $\mathcal{CN}(a_{nn_\text{R}};\hat{p}_{nn_\text{R}}^{u_\text{I}},v^{u_\text{I}}_{p,nn_\text{R}})$, where
	\begin{subequations}\label{equ_pv}
	\begin{align}
		\hat{p}^{u_\text{I}}_{nn_\text{R}}&=\sum_{k=1}^{K}\hat{b}^{u_\text{I}}_{nk}\hat{g}^{u_\text{I}}_{kn_\text{R}}-\hat{\beta}^{u_\text{I}-1}_{nn_\text{R}}(|\hat{b}^{u_\text{I}}_{nk}|^2v_{g,kn_\text{R}}^{u_\text{I}}+v_{b,nk}^{u_\text{I}}|\hat{g}^{u_\text{I}}_{kn_\text{R}}|^2)
		\\
		v_{p,nn_\text{R}}^{u_\text{I}}&=\sum_{k=1}^{K}|\hat{b}^{u_\text{I}}_{nk}|^2v_{g,kn_\text{R}}^{u_\text{I}}+v_{b,nk}^{u_\text{I}}|\hat{g}^{u_\text{I}}_{kn_\text{R}}|^2+v_{b,nk}^{u_\text{I}}v_{g,kn_\text{R}}^{u_\text{I}}.
	\end{align}
	\end{subequations}
	The message from node  $p(a_{nn_\text{R}}|\cdot)$ to node $g_{kn_\text{R}}$ at iteration $u_\text{I}$, denoted by $\mathcal{M}^{u_\text{I}}_{p(a_{nn_\text{R}}|\cdot)\rightarrow g_{kn_\text{R}}}(g_{kn_\text{R}})$, is approximated by $\mathcal{CN}(g_{kn_\text{R}};\hat{q}_{kn_\text{R}}^{u_\text{I}},v^{u_\text{I}}_{q,kn_\text{R}})$, where
	\begin{subequations}\label{equ_qaap}
		\begin{align}
			\nonumber\hat{q}_{kn_\text{R}}^{u_\text{I}} &= \hat{g}_{kn_\text{R}}^{u_\text{I}}(1-v^{u_\text{I}}_{q,kn_\text{R}}\sum_{n=1}^{N_\text{W}M_{\text{OSF}}}v^{u_\text{I}}_{b,nk}v^{u_\text{I}}_{\beta,nn_\text{R}})\\&\hspace{0.5cm}+v^{u_\text{I}}_{q,kn_\text{R}}\sum_{n=1}^{N_\text{W}M_{\text{OSF}}}(\hat{b}^{u_\text{I}}_{nk})^*\hat{\beta}^{u_\text{I}}_{nn_\text{R}}
			\\
			v^{u_\text{I}}_{q,kn_\text{R}} &= (\sum_{n=1}^{N_\text{W}M_{\text{OSF}}}(\hat{b}^{u_\text{I}}_{nk})^2 v^{u_\text{I}}_{\beta,nn_\text{R}})^{-1}
		\end{align}
	\end{subequations}	
	with
	\begin{equation}\nonumber
		\hat{\beta}^{u_\text{I}}_{nn_\text{R}}=\frac{\hat{a}^{u_\text{I}}_{nn_\text{R}}-\hat{p}^{u_\text{I}}_{nn_\text{R}}}{v_{p,nn_\text{R}}^{u_\text{I}}}
		\quad
		v^{u_\text{I}}_{\beta,nn_\text{R}}=\left(\frac{1-v_{a,nn_\text{R}}^{u_\text{I}}}{v_{p,nn_\text{R}}^{u_\text{I}}}\right)\frac{1}{v_{p,nn_\text{R}}^{u_\text{I}}}.
	\end{equation}
	Herein, $\hat{a}^{u_\text{I}}_{nn_\text{R}}$ and $v^{u_\text{I}}_{a,nn_\text{R}}$ are the marginal posterior mean and variance of $a_{nn_\text{R}}$ at iteration $u_\text{I}$, respectively.
	
	Similarly, the message at iteration $u_\text{I}$ from node $p(a_{nn_\text{R}}|\cdot)$ to node $b_{nk}$, denoted by $\mathcal{M}^{u_\text{I}}_{p(a_{nn_\text{R}}|\cdot)\rightarrow b_{nk}}(b_{nk})$, is approximated by $\mathcal{CN}(b_{nk};\hat{r}_{nk}^{u_\text{I}},v^{u_\text{I}}_{r,nk})$, where
	\begin{subequations}\label{equ_rapp}
		\begin{align}
			\nonumber&\hat{r}_{nk}^{u_\text{I}} = \hat{b}_{nk}^{u_\text{I}}(1-v^{u_\text{I}}_{r,nk}\sum_{n_\text{R}=1}^{N_\text{R}}v^{u_\text{I}}_{\beta,nn_\text{R}}v^{u_\text{I}}_{g,kn_\text{R}})\\&\hspace{1cm}+v^{u_\text{I}}_{r,nk}\sum_{n_\text{R}=1}^{N_\text{R}}\hat{\beta}^{u_\text{I}}_{nn_\text{R}}(\hat{g}^{u_\text{I}}_{kn_\text{R}})^*
			\\
			&v^{u_\text{I}}_{r,nk} = (\sum_{n_\text{R}=1}^{N_\text{R}}v^{u_\text{I}}_{\beta,nn_\text{R}}(\hat{g}^{u_\text{I}}_{kn_\text{R}})^2)^{-1}.
		\end{align}
	\end{subequations}
	Based on the linear model $b_{nk}=\sum_{i=1}^{N_\text{W}M_{\text{OSF}}}z_{ni}x_{ik}$, the message at iteration $u_\text{I}$ from node  $p(b_{nk}|\cdot)$ to node $b_{nk}$, denoted by $\mathcal{M}^{u_\text{I}}_{p(b_{nk}|\cdot)\rightarrow b_{nk}}(b_{nk})$, is approximated by $\mathcal{CN}(b_{nk};\hat{o}_{nk}^{u_\text{I}},v^{u_\text{I}}_{o,nk})$, where
	\begin{subequations}\label{equ_oapp}
		\begin{align}
			\hat{o}^{u_\text{I}}_{nk}&=\sum_{i=1}^{N_\text{W}M_{\text{OSF}}}z_{ni}\hat{x}^{u_\text{I}}_{ik}-\hat{\gamma}^{u_\text{I}-1}_{nk}\sum_{i=1}^{N_\text{W}M_{\text{OSF}}}(z_{ni})^2v_{x,ik}^{u_\text{I}}
			\\
			v_{o,nk}^{u_\text{I}}&=\sum_{i=1}^{N_\text{W}M_{\text{OSF}}}(z_{ni})^2v_{x,ik}^{u_\text{I}}.
		\end{align}
	\end{subequations}
	The message from node $p(b_{nk}|\cdot)$ to node $x_{ik}$ at iteration $u_\text{I}$, denoted by $\mathcal{M}^{u_\text{I}}_{p(b_{nk}|\cdot)\rightarrow x_{ik}}(x_{ik})$, is approximated by $\mathcal{CN}(x_{ik};\hat{m}_{ik}^{u_\text{I}},v^{u_\text{I}}_{m,ik})$, where 	\begin{subequations}\label{equ_mapp}
		\begin{align}
			\hat{m}_{ik}^{u_\text{I}} &= \hat{x}_{ik}^{u_\text{I}}+v^{u_\text{I}}_{m,ik}\sum_{n=1}^{N_\text{W}M_{\text{OSF}}}(z_{ni})^*\hat{\gamma}^{u_\text{I}}_{nk}
			\\
			v^{u_\text{I}}_{m,ik} &= \left(\sum_{n=1}^{N_\text{W}M_{\text{OSF}}}(z_{ni})^2v^{u_\text{I}}_{\gamma,nk}\right)^{-1}
		\end{align}
	\end{subequations}
	with
	\begin{equation}\nonumber
		\hat{\gamma}^{u_\text{I}}_{nk}=\frac{\hat{b}^{u_\text{I}+1}_{nk}-\hat{o}^{u_\text{I}}_{nk}}{v_{o,nk}^{u_\text{I}}}\quad			
		v^{u_\text{I}}_{\gamma,nk}=\left(\frac{1-v_{b,nk}^{u_\text{I}+1}}{v_{o,nk}^{u_\text{I}}}\right)\frac{1}{v_{o,nk}^{u_\text{I}}}.
	\end{equation}
	Herein, $\hat{b}^{u_\text{I}+1}_{nk}$ and $v^{u_\text{I}+1}_{b,nk}$ are the marginal posterior mean and variance of $b_{nk}$ at iteration $u_\text{I}+1$, respectively.
	
	The proposed JUCED algorithm is summarized in Algorithm \ref{alg_jcedd}, where $\varepsilon_1$ is a predefined small value, $\{\hat{x}^{u_\text{O}}_{ik}\}$ and $\{v^{u_\text{O}}_{x,ik}\}$ are the posterior mean and variance of $p(\mathbf{X}|\mathbf{Y};\hat{\bm{\tau}}^{u_\text{O}})$, respectively, $\{\hat{g}_{kn_\text{R}}^{u_\text{O}}\}$ and $\{v^{u_\text{O}}_{g,kn_\text{R}}\}$ are the posterior mean and variance of $p(\mathbf{G}|\mathbf{Y};\hat{\bm{\tau}}^{u_\text{O}})$, respectively. From \eqref{equ_mmse}, the MMSE estimate of $\mathbf{X}$ and $\mathbf{G}$ in the $u_\text{O}$-th iteration are  $\{\hat{x}^{u_\text{O}}_{ik}\}$ and $\{\hat{g}^{u_\text{O}}_{kn_\text{R}}\}$, respectively. The $k$-th user is active if the power of the $k$-th row of $\{\hat{g}^{u_\text{O}}_{kn_\text{R}}\}$ is larger than a predetermined threshold $\eta_\text{th}$, i.e.,
	\begin{equation}\label{equ_active}
		\hat{\alpha}_k^{u_\text{O}}=\begin{cases}
			1,\quad \sum_{n_\text{R}=1}^{N_\text{R}}|\hat{g}^{u_\text{O}}_{kn_\text{R}}|^2>\eta_\text{th},\\
			0,\quad \sum_{n_\text{R}=1}^{N_\text{R}}|\hat{g}^{u_\text{O}}_{kn_\text{R}}|^2\leq\eta_\text{th}.
		\end{cases}
	\end{equation}
	\begin{algorithm}[!htbp] 
		\caption{JUCED}
		\begin{algorithmic}[1] 
			\STATEx \textbf{Input: }$\mathbf{Y}$, $\mathbf{Z}$, $\lambda$, $\sigma_\text{W}^2$, $\hat{\bm{\tau}}^{u_\text{O}}$,  $\rho$
			\FOR{$u_\text{I}=1:U_\text{I}$}  
			\State $\forall n,n_\text{R}:$ compute  $\hat{p}^{u_\text{I}}_{nn_\text{R}}$ and $v_{p,nn_\text{R}}^{u_\text{I}}$ by \eqref{equ_pv};
			\State $\forall n,n_\text{R}:$ compute  $\hat{a}^{u_\text{I}}_{nn_\text{R}}=\text{E}\{a_{nn_\text{R}}|\hat{p}^{u_\text{I}}_{nn_\text{R}},v_{p,nn_\text{R}}^{u_\text{I}}\}$ and $v^{u_\text{I}}_{a,nn_\text{R}}=\text{Var}\{a_{nn_\text{R}}|\hat{p}^{u_\text{I}}_{nn_\text{R}},v_{p,nn_\text{R}}^{u_\text{I}}\}$;
			\State $\forall k,n_\text{R}:$ compute  $\hat{q}_{kn_\text{R}}^{u_\text{I}}$ and $v^{u_\text{I}}_{q,kn_\text{R}}$ by \eqref{equ_qaap};
			\State $\forall k,n_\text{R}:$ compute  $\hat{g}^{u_\text{I}+1}_{kn_\text{R}}=\text{E}\{g_{kn_\text{R}}|\hat{q}_{kn_\text{R}}^{u_\text{I}},v^{u_\text{I}}_{q,kn_\text{R}}\}$ and $v^{u_\text{I}+1}_{g,kn_\text{R}}=\text{Var}\{g_{kn_\text{R}}|\hat{q}_{kn_\text{R}}^{u_\text{I}},v^{u_\text{I}}_{q,kn_\text{R}}\}$;
			\State $\forall n,k:$ compute  $\hat{r}_{nk}^{u_\text{I}}$ and $v^{u_\text{I}}_{r,nk}$ by \eqref{equ_rapp};
			\State $\forall n,k:$ compute  $\hat{o}^{u_\text{I}}_{nk}$ and $v_{o,nk}^{u_\text{I}}$ by \eqref{equ_oapp};
			\State $\forall n,k:$ compute  $\hat{b}^{u_\text{I}+1}_{nk}=\text{E}\{b_{nk}|\hat{r}_{nk}^{u_\text{I}},v^{u_\text{I}}_{r,nk},\hat{o}_{nk}^{u_\text{I}},v^{u_\text{I}}_{o,nk}\}$ and $v^{u_\text{I}+1}_{b,nk}=\text{Var}\{b_{nk}|\hat{r}_{nk}^{u_\text{I}},v^{u_\text{I}}_{r,nk},\hat{o}_{nk}^{u_\text{I}},v^{u_\text{I}}_{o,nk}\}$;
			\State $\forall i,k:$ compute  $\hat{m}_{ik}^{u_\text{I}}$ and $v^{u_\text{I}}_{m,ik}$ by \eqref{equ_mapp};
			\State $\forall i,k:$ compute  $\hat{x}^{u_\text{I}+1}_{ik}=\text{E}\{x_{ik}|\hat{m}_{ik}^{u_\text{I}},v^{u_\text{I}}_{m,ik}\}$ and $v^{u_\text{I}+1}_{x,ik}=\text{Var}\{x_{ik}|\hat{m}_{ik}^{u_\text{I}},v^{u_\text{I}}_{m,ik}\}$;
			\State $\forall k:$ identify active users by \eqref{equ_active};
			\State \textbf{if}  $\sum_{k}\sum_{n_\text{R}}|v^{u_\text{I}+1}_{g,kn_\text{R}}-v^{u_\text{I}}_{g,kn_\text{R}}|<\varepsilon_1$ \textbf{stop}
			\ENDFOR
			\STATEx \textbf{Output: }$\{\hat{x}^{u_\text{O}}_{ik},v^{u_\text{O}}_{x,ik}\}$,  $\{\hat{g}^{u_\text{O}}_{kn_\text{R}},v^{u_\text{O}}_{g,kn_\text{R}}\}$
		\end{algorithmic} 
		\label{alg_jcedd}
	\end{algorithm} 
	
	
	\subsection{Overall Algorithm}\label{subsec_overall}
	The proposed delay-calibrated JUCED algorithm is summarized in Algorithm \ref{alg_jdcuad}, where $\varepsilon_2$ is a predefined small value. 
	
	\begin{algorithm}[!htbp] 
		\caption{Delay-Calibrated JUCED}
		\begin{algorithmic}[1] 
			\STATEx \textbf{Input: }$\mathbf{Y}$, $\mathbf{Z}$, $\lambda$, $\sigma_\text{W}^2$, $\rho$
			\STATEx \% Initialization
			\STATE Calculate cross-correlations with local preamble sequences, find the peaks and store their time delays as $\hat{\bm{\tau}}^{1}$; 
			\STATE Apply the pre-whitening approach;
			\STATEx \% Main part
			\FOR{$u_\text{O}=1:U_\text{O}$} 
			\STATE Use Algorithm \ref{alg_jcedd} to obtain $\{\hat{x}^{u_\text{O}}_{ik},v^{u_\text{O}}_{x,ik}\}$,  $\{\hat{g}^{u_\text{O}}_{kn_\text{R}},v^{u_\text{O}}_{g,kn_\text{R}}\}$;
			\STATE Use the greedy search based delay-calibration algorithm \cite{shao2024} to obtain $\hat{\bm{\tau}}^{u_\text{O}+1}$;
			\State \textbf{if} $\sum_{k}|\hat{\tau}_k^{u_\text{O}+1}-\hat{\tau}_k^{u_\text{O}}|<\varepsilon_2$ \textbf{stop}
			\ENDFOR
			\STATEx \textbf{Output: }$\hat{\bm{\tau}}$, $\{\hat{x}_{ik}\}$,  $\{\hat{g}_{kn_\text{R}}\}$
		\end{algorithmic} 
		\label{alg_jdcuad}
	\end{algorithm} 
	
\section{Numerical Results}\label{numer}
We evaluate the proposed JUCED algorithm using simulations and the setup is as follows. The root-raised-cosine filter is used for the pulse shaping filter. The AP is equipped with $N_\text{R}=64$ receive antennas and serves $300$ single-antenna devices. The probability of active users is 0.25. There are $64$ available orthogonal preamble sequences in the service region. The Zadoff–Chu (ZC) sequences are adopted for the generation of preambles due to its good auto-correlation properties and low peak-to-average power ratio \cite{4119357}. Moreover, $\lambda$ in dB is set as $(-128.1-118.1)/2$. The probability of false alarm is set to lower than $10^{-3}$. 

\begin{figure}[!htbp]
	\begin{minipage}{\columnwidth}
		\centering
		\begin{tikzpicture} 
	\begin{axis}[%
		hide axis,
		xmin=10,
		xmax=40,
		ymin=0,
		ymax=0.4,
		legend columns=2,
		legend style={draw=white!15!black,legend cell align=left,font=\footnotesize}
		]
		
		\addlegendimage{color=red,dashed,line width=0.8pt,mark=star,mark options={solid},mark size=3pt}
		\addlegendentry{UAD-DC \cite{shao2024} w/ $M_{\text{OSF}}=1$};
		
		\addlegendimage{color=blue,dashed,line width=0.8pt,mark=triangle,mark options={solid},mark size=3pt}
		\addlegendentry{UAD-DC \cite{shao2024} w/ $M_{\text{OSF}}=2$};
		
		\addlegendimage{color=black,dashed,line width=0.8pt,mark=square,mark options={solid},mark size=3pt}
		\addlegendentry{UAD-DC \cite{shao2024} w/ $M_{\text{OSF}}=3$};
		
		\addlegendimage{color=red,solid,line width=0.8pt,mark=star,mark options={solid},mark size=3pt}
		\addlegendentry{Algorithm \ref{alg_jdcuad} w/ $M_{\text{OSF}}=1$};
		
		\addlegendimage{color=blue,solid,line width=0.8pt,mark=triangle,mark options={solid},mark size=3pt}
		\addlegendentry{Algorithm \ref{alg_jdcuad} w/ $M_{\text{OSF}}=2$};
		
		\addlegendimage{color=black,solid,line width=0.8pt,mark=square,mark options={solid},mark size=3pt}
		\addlegendentry{Algorithm \ref{alg_jdcuad} w/ $M_{\text{OSF}}=3$};
		
	\end{axis}
\end{tikzpicture}
	\end{minipage}\vfil
	\begin{minipage}{\columnwidth}
		\begin{minipage}{0.43\columnwidth}
			\begin{tikzpicture}
	
	\begin{axis}[%
		width=0.8\columnwidth,
		height=\columnwidth,
		at={(0in,0in)},
		scale only axis,
		xmin=0,
		xmax=20,
		xlabel={SNR},
		xtick=data,
		xmajorgrids,
		ymode=log,
		ymin=1e-1,
		ymax=10,
		yminorticks=true,
		label style={font=\scriptsize},
		y label style={at={(axis description cs:-0.15,.5)}},
		tick label style={font=\scriptsize},
		ylabel={NMSE of $\mathbf{X}$},
		axis background/.style={fill=white},
		ymajorgrids,
		yminorgrids,
		legend style={at={(0.45,0.75)},anchor=south west,legend cell align=left,align=left,draw=white!15!black,font=\scriptsize}
		]
		
		\addplot [color=red, solid,line width=1pt,mark=star,mark options={solid},mark size=3pt]
		table[row sep=crcr]{%
			0     0.8116\\
			5     0.6974\\
			10    0.6163\\
			15    0.6049\\
			20    0.6374\\
		};
		
		\addplot [color=blue, solid,line width=1pt,mark=triangle,mark options={solid},mark size=3pt]
		table[row sep=crcr]{%
			0     0.7800\\
			5     0.6315\\
			10    0.4505\\
			15    0.3117\\
			20    0.2490\\
		};
		
		\addplot [color=black, solid,line width=1pt,mark=square,mark options={solid},mark size=3pt]
		table[row sep=crcr]{%
			0     0.7504\\
			5     0.5834\\
			10    0.4102\\
			15    0.2715\\
			20    0.2053\\
		};
		
		\addplot [color=red, dashed,line width=1pt,mark=star,mark options={solid},mark size=3pt]
		table[row sep=crcr]{%
			0     1.9629\\
			5     1.2556\\
			10    1.0480\\
			15    0.9427\\
			20    0.9404\\
		};
		
		\addplot [color=blue, dashed,line width=1pt,mark=triangle,mark options={solid},mark size=3pt]
		table[row sep=crcr]{%
			0     1.8551\\
			5     1.2337\\
			10    1.0514\\
			15    0.9631\\
			20    0.9189\\
		};
		
		\addplot [color=black, dashed,line width=1pt,mark=square,mark options={solid},mark size=3pt]
		table[row sep=crcr]{%
			0     1.7684\\
			5     1.3712\\
			10    1.0243\\
			15    0.8677\\
			20    0.7934\\
		};
		
	\end{axis}
\end{tikzpicture}
			\label{fig_ber}
		\end{minipage}\hfil
		\begin{minipage}{0.43\columnwidth}
			\begin{tikzpicture}
	
	\begin{axis}[%
		width=0.8\columnwidth,
		height=\columnwidth,
		at={(0in,0in)},
		scale only axis,
		xmin=0,
		xmax=20,
		xlabel={SNR},
		xtick=data,
		xmajorgrids,
		ymin=0,
		ymax=1,
		yminorticks=true,
		label style={font=\scriptsize},
		y label style={at={(axis description cs:-0.15,.5)}},
		tick label style={font=\scriptsize},
		ylabel={Probability of misdetection},
		axis background/.style={fill=white},
		ymajorgrids,
		yminorgrids,
		legend style={at={(0.45,0.75)},anchor=south west,legend cell align=left,align=left,draw=white!15!black,font=\scriptsize}
		]
		
		\addplot [color=red, solid,line width=1pt,mark=star,mark options={solid},mark size=3pt]
		table[row sep=crcr]{%
			0     0.6648\\
			5     0.4818\\
			10    0.4267\\
			15    0.4278\\
			20    0.4204\\
		};
		
		\addplot [color=blue, solid,line width=1pt,mark=triangle,mark options={solid},mark size=3pt]
		table[row sep=crcr]{%
			0     0.5726\\
			5     0.3190\\
			10    0.245\\
			15    0.2244\\
			20    0.2262\\
		};
		
		\addplot [color=black, solid,line width=1pt,mark=square,mark options={solid},mark size=3pt]
		table[row sep=crcr]{%
			0     0.4671\\
			5     0.2836\\
			10    0.2204\\
			15    0.1996\\
			20    0.1938\\
		};
		
		\addplot [color=red, dashed,line width=1pt,mark=star,mark options={solid},mark size=3pt]
		table[row sep=crcr]{%
			0     0.8200\\
			5     0.7227\\
			10    0.6653\\
			15    0.6640\\
			20    0.6520\\
		};
		
		\addplot [color=blue, dashed,line width=1pt,mark=triangle,mark options={solid},mark size=3pt]
		table[row sep=crcr]{%
			0     0.7920\\
			5     0.6760\\
			10    0.6347\\
			15    0.6160\\
			20    0.6040\\
		};
		
		\addplot [color=black, dashed,line width=1pt,mark=square,mark options={solid},mark size=3pt]
		table[row sep=crcr]{%
			0     0.7540\\
			5     0.6220\\
			10    0.5587\\
			15    0.5360\\
			20    0.5247\\
		};
		
	\end{axis}
\end{tikzpicture}
			\label{fig_pmd2}
		\end{minipage}
	\end{minipage}
	\vspace{-0.5cm}
	\caption{Performance comparisons under different SNRs.}\label{fig_peralg}
\end{figure}
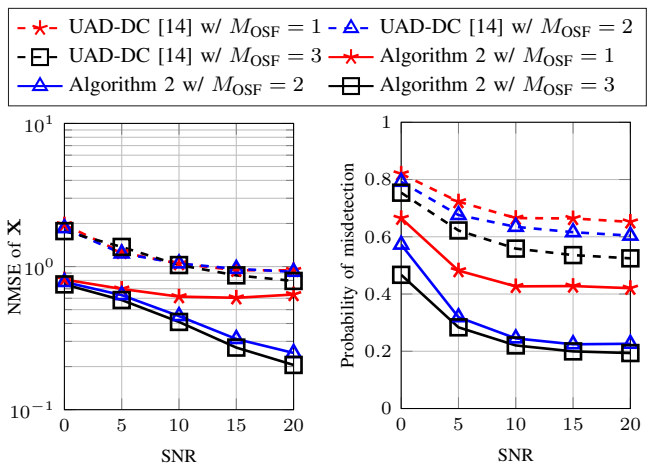

Fig. \ref{fig_peralg} shows the NMSEs of $\mathbf{G}$ and $\mathbf{X}$, and the probability of misdetection against different SNRs with $N_\text{P}=67$, $N_\text{D}=128$. It can be seen that oversampling can significantly improve the performance, as accurate estimation of time delays helps to detect more users shared with the same preamble, estimate the channels more precisely, and recover data more accurately. Moreover, we have compared the performance of Algorithm \ref{alg_jdcuad} with that of UAD-DC proposed in \cite{shao2024}. Since UAD-DC \cite{shao2024} only solves the channel estimation problem, we employ the least squares estimator to detect data symbols after obtaining the estimated channel matrix. Note that the NMSEs of $\mathbf{G}$ increase with larger SNR since UAD-DC \cite{shao2024} does not consider the interference from the data, which dominates at high SNRs. It can be seen that Algorithm \ref{alg_jdcuad} significantly outperforms UAD-DC \cite{shao2024}. For example, when SNR = 20 dB and $M_{\text{OSF}}=2$, the NMSEs of $\mathbf{G}$ and $\mathbf{X}$ of Algorithm \ref{alg_jdcuad} are reduced by 80.36\% and 72.90\%, respectively, in comparison to those of UAD-DC \cite{shao2024}.

\section{Conclusions}
In this work, we proposed the delay-calibrated JUCED algorithm for asynchronous mMTC. The time delays of active users were iteratively calibrated by the expectation-maximization method. Based on the calibrated time delays, a message passing based JUCED algorithm was proposed. Numerical results demonstrate the effectiveness of the proposed algorithms in terms of the NMSEs of channel and data symbols, and the probability of misdetection. Compared to the baseline method, the proposed JUCED algorithm can achieve performance improvements of over 70\%.

\section{Acknowledgment}
This work was supported in part by the Research start-up funds under Grant No. 2025K001, CAPES, CNPq, FAPERJ, and FAPESP.

\bibliographystyle{IEEEtran}
\bibliography{ref}

\begin{thebibliography}{10}
\providecommand{\url}[1]{#1}
\csname url@samestyle\endcsname
\providecommand{\newblock}{\relax}
\providecommand{\bibinfo}[2]{#2}
\providecommand{\BIBentrySTDinterwordspacing}{\spaceskip=0pt\relax}
\providecommand{\BIBentryALTinterwordstretchfactor}{4}
\providecommand{\BIBentryALTinterwordspacing}{\spaceskip=\fontdimen2\font plus
\BIBentryALTinterwordstretchfactor\fontdimen3\font minus \fontdimen4\font\relax}
\providecommand{\BIBforeignlanguage}[2]{{%
\expandafter\ifx\csname l@#1\endcsname\relax
\typeout{** WARNING: IEEEtran.bst: No hyphenation pattern has been}%
\typeout{** loaded for the language `#1'. Using the pattern for}%
\typeout{** the default language instead.}%
\else
\language=\csname l@#1\endcsname
\fi
#2}}
\providecommand{\BIBdecl}{\relax}
\BIBdecl

\bibitem{9205230}
X.~Chen, D.~W.~K. Ng, W.~Yu, E.~G. Larsson, N.~Al-Dhahir, and R.~Schober, ``Massive access for {5G} and beyond,'' \emph{{IEEE} J. Sel. Areas Commun.}, vol.~39, no.~3, pp. 615--637, 2021.

\bibitem{detmtc}
R.~B. Di~Renna, C.~Bockelmann, R.~C. de~Lamare, and A.~Dekorsy, ``Detection techniques for massive machine-type communications: Challenges and solutions,'' \emph{IEEE Access}, vol.~8, pp. 180\,928--180\,954, 2020.

\bibitem{6525600}
M.~Hasan, E.~Hossain, and D.~Niyato, ``Random access for machine-to-machine communication in {LTE}-advanced networks: issues and approaches,'' \emph{{IEEE} Commun. Mag.}, vol.~51, no.~6, pp. 86--93, 2013.

\bibitem{7565189}
C.~Bockelmann, N.~Pratas, H.~Nikopour, K.~Au, T.~Svensson, C.~Stefanovic, P.~Popovski, and A.~Dekorsy, ``Massive machine-type communications in 5{G}: physical and {MAC}-layer solutions,'' \emph{{IEEE} Commun. Mag.}, vol.~54, no.~9, pp. 59--65, 2016.

\bibitem{9537931}
J.~Choi, J.~Ding, N.-P. Le, and Z.~Ding, ``Grant-free random access in machine-type communication: Approaches and challenges,'' \emph{{IEEE} Wireless Commun.}, vol.~29, no.~1, pp. 151--158, 2022.

\bibitem{8454392}
L.~Liu, E.~G. Larsson, W.~Yu, P.~Popovski, C.~Stefanovic, and E.~de~Carvalho, ``Sparse signal processing for grant-free massive connectivity: A future paradigm for random access protocols in the internet of things,'' \emph{{IEEE} Signal Process. Mag.}, vol.~35, no.~5, pp. 88--99, 2018.

\bibitem{listmtc}
R.~B. Di~Renna and R.~C. de~Lamare, ``Iterative list detection and decoding for massive machine-type communications,'' \emph{IEEE Transactions on Communications}, vol.~68, no.~10, pp. 6276--6288, 2020.

\bibitem{9140386}
S.~Jiang, X.~Yuan, X.~Wang, C.~Xu, and W.~Yu, ``Joint user identification, channel estimation, and signal detection for grant-free noma,'' \emph{{IEEE} Trans. Wireless Commun.}, vol.~19, no.~10, pp. 6960--6976, 2020.

\bibitem{9103622}
Y.~Zhang, Z.~Yuan, Q.~Guo, Z.~Wang, J.~Xi, and Y.~Li, ``Bayesian receiver design for grant-free noma with message passing based structured signal estimation,'' \emph{{IEEE} Trans. Veh. Technol.}, vol.~69, no.~8, pp. 8643--8656, 2020.

\bibitem{9714456}
R.~B. Di~Renna and R.~C. de~Lamare, ``Joint channel estimation, activity detection and data decoding based on dynamic message-scheduling strategies for mmtc,'' \emph{{IEEE} Trans. Commun.}, vol.~70, no.~4, pp. 2464--2479, 2022.

\bibitem{9691883}
Z.~Wang, Y.-F. Liu, and L.~Liu, ``Covariance-based joint device activity and delay detection in asynchronous mmtc,'' \emph{{IEEE} Signal Process. Lett.}, vol.~29, pp. 538--542, 2022.

\bibitem{9390399}
W.~Zhu, M.~Tao, X.~Yuan, and Y.~Guan, ``Deep-learned approximate message passing for asynchronous massive connectivity,'' \emph{{IEEE} Trans. Wireless Commun.}, vol.~20, no.~8, pp. 5434--5448, 2021.

\bibitem{msgamp}
R.~B.~D. Renna and R.~C. de~Lamare, ``Dynamic message scheduling based on activity-aware residual belief propagation for asynchronous mmtc,'' \emph{IEEE Wireless Communications Letters}, vol.~10, no.~6, pp. 1290--1294, 2021.

\bibitem{shao2024}
\BIBentryALTinterwordspacing
Z.~Shao, X.~Yuan, R.~C. de~Lamare, and Y.~Zhang, ``User activity detection with delay-calibration for asynchronous massive random access,'' 2024. [Online]. Available: \url{https://arxiv.org/abs/2411.01923}
\BIBentrySTDinterwordspacing

\bibitem{jidf}
R.~C. de~Lamare and R.~Sampaio-Neto, ``Adaptive reduced-rank processing based on joint and iterative interpolation, decimation, and filtering,'' \emph{IEEE Transactions on Signal Processing}, vol.~57, no.~7, pp. 2503--2514, 2009.

\bibitem{dynovs}
Z.~Shao, L.~T.~N. Landau, and R.~C. de~Lamare, ``Dynamic oversampling for 1-bit adcs in large-scale multiple-antenna systems,'' \emph{IEEE Transactions on Communications}, vol.~69, no.~5, pp. 3423--3435, 2021.

\bibitem{em}
A.~P. Dempster, N.~M. Laird, and D.~B. Rubin, ``Maximum likelihood from incomplete data via the {EM} algorithm,'' \emph{Journal of the Royal Statistical Society: Series B (Methodological)}, vol.~39, no.~1, pp. 1--22, 1977.

\bibitem{spa}
R.~C. De~Lamare and R.~Sampaio-Neto, ``Minimum mean-squared error iterative successive parallel arbitrated decision feedback detectors for ds-cdma systems,'' \emph{IEEE Transactions on Communications}, vol.~56, no.~5, pp. 778--789, 2008.

\bibitem{mfsic}
P.~Li, R.~C. de~Lamare, and R.~Fa, ``Multiple feedback successive interference cancellation detection for multiuser mimo systems,'' \emph{IEEE Transactions on Wireless Communications}, vol.~10, no.~8, pp. 2434--2439, 2011.

\bibitem{mbdf}
R.~C. de~Lamare, ``Adaptive and iterative multi-branch mmse decision feedback detection algorithms for multi-antenna systems,'' \emph{IEEE Transactions on Wireless Communications}, vol.~12, no.~10, pp. 5294--5308, 2013.

\bibitem{bfidd}
A.~G.~D. Uchoa, C.~T. Healy, and R.~C. de~Lamare, ``Iterative detection and decoding algorithms for mimo systems in block-fading channels using ldpc codes,'' \emph{IEEE Transactions on Vehicular Technology}, vol.~65, no.~4, pp. 2735--2741, 2016.

\bibitem{1bitidd}
Z.~Shao, R.~C. de~Lamare, and L.~T.~N. Landau, ``Iterative detection and decoding for large-scale multiple-antenna systems with 1-bit adcs,'' \emph{IEEE Wireless Communications Letters}, vol.~7, no.~3, pp. 476--479, 2018.

\bibitem{llrref}
T.~Ssettumba, Z.~Shao, L.~T.~N. Landau, M.~Soares Pereira~Facina, P.~Branco~da Silva, and R.~C. de~Lamare, ``Centralized and decentralized idd schemes for cell-free massive mimo systems: Ap selection and llr refinement,'' \emph{IEEE Access}, vol.~12, pp. 62\,392--62\,406, 2024.

\bibitem{risidd}
R.~C.~G. Porto and R.~C. de~Lamare, ``Iterative detection and decoding for multiuser systems based on mmse refinements with active or passive ris,'' \emph{IEEE Wireless Communications Letters}, vol.~14, no.~1, pp. 208--212, 2025.

\bibitem{apsllr}
R.~B.~D. Renna and R.~C. de~Lamare, ``Iterative detection and decoding with log-likelihood ratio based access point selection for cell-free mimo systems,'' \emph{IEEE Transactions on Vehicular Technology}, vol.~73, no.~5, pp. 7418--7423, 2024.

\bibitem{iddocl}
T.~Ssettumba, S.~Mashdour, L.~T.~N. Landau, P.~da~Silva, and R.~C. de~Lamare, ``Iterative interference cancellation for clustered cell-free massive mimo networks,'' \emph{IEEE Wireless Communications Letters}, vol.~14, no.~2, pp. 509--513, 2025.

\bibitem{4119357}
Y.~Wen, W.~Huang, and Z.~Zhang, ``{CAZAC} sequence and its application in {LTE} random access,'' in \emph{IEEE Information Theory Workshop}, 2006, pp. 544--547.

\end{thebibliography}

\end{document}